\documentclass[conference,twoside]{IEEEtran}
\IEEEoverridecommandlockouts

\usepackage{dcolumn}
\usepackage{longtable}
\usepackage{cite}
\usepackage{amsmath,amssymb,amsfonts}
\usepackage{algorithmic}
\usepackage{graphicx}
\usepackage{textcomp}
\usepackage{xcolor}
\usepackage{multicol}
\usepackage{orcidlink}
\usepackage{hyperref}
\usepackage{url}

\def\BibTeX{{\rm B\kern-.05em{\sc i\kern-.025em b}\kern-.08em
    T\kern-.1667em\lower.7ex\hbox{E}\kern-.125emX}}

\begin{document}

\title{A Systematic Review of Common Beginner Programming Mistakes in Data Engineering}

\author{\IEEEauthorblockN{Max Neuwinger}
\IEEEauthorblockA{\textit{Professorship for Open-Source Software} \\
\textit{Friedrich-Alexander University Erlangen-Nürnberg}\\
Erlangen, Germany \\
max.neuwinger@fau.de \orcidlink{0009-0003-3930-700X}}
\and
\IEEEauthorblockN{Dirk Riehle}
\IEEEauthorblockA{\textit{Professorship for Open-Source Software} \\
\textit{Friedrich-Alexander University Erlangen-Nürnberg}\\
Erlangen, Germany \\
dirk@riehle.org \orcidlink{0000-0002-8139-5600}}
}

\maketitle

\begin{abstract}
The design of effective programming languages, libraries, frameworks, tools, and platforms for data engineering strongly depends on their ease and correctness of use. Anyone who ignores that it is humans who use these tools risks building tools that are useless, or worse, harmful. To ensure our data engineering tools are based on solid foundations, we performed a systematic review of common programming mistakes in data engineering. We focus on programming beginners (students) by analyzing both the limited literature specific to data engineering mistakes and general programming mistakes in languages commonly used in data engineering (Python, SQL, Java). Through analysis of 21 publications spanning from 2003 to 2024, we synthesized these complementary sources into a comprehensive classification that captures both general programming challenges and domain-specific data engineering mistakes. This classification provides an empirical foundation for future tool development and educational strategies. We believe our systematic categorization will help researchers, practitioners, and educators better understand and address the challenges faced by novice data engineers.
\end{abstract}


\section{Introduction}

The rapid growth of data science and artificial intelligence applications has created strong demand for skilled data engineers. According to the U.S. Bureau of Labor Statistics, data scientist positions are projected to grow by 35\% from 2022 to 2032 \cite{bls2024data}, while the World Economic Forum expects similar growth of 30-35\% in demand for data analysts, scientists, and engineers by 2027 \cite{wef2023future}.

As the field grows, more beginners and students are entering the profession, leading to wider ranges of skill levels among practitioners. The essential role of data engineering in the data science pipeline makes this variation in expertise particularly important. Data scientists reportedly spend up to 80\% of their time resolving data issues upstream of modeling activities \cite{howe2017data}, showing how programming mistakes can significantly affect data integrity, analysis accuracy, and decision-making processes.

The challenges are especially clear in educational settings, where research shows that students struggle with abstract programming concepts and algorithm design \cite{konecki2014problems}. This mix of growing demand, varying skill levels, and learning challenges highlights the need for better tools and educational practices. Our research addresses this need by systematically identifying common beginner programming mistakes in data engineering, aiming to develop more effective educational methods.

This work was motivated by our research into domain-specific languages (DSL) for data engineering (the Jayvee\footnote{See \url{https://github.com/jvalue/jayvee}} DSL of the JValue\footnote{See \url{https://jvalue.com}} research project). To ground our design decisions in empirical evidence and educational insights, we conducted a systematic review of existing literature on common programming mistakes in data engineering, as presented in this article. This review informs both our tool development process and the formulation of educational recommendations, which we are currently validating and will continue to evaluate through controlled experiments assessing the effectiveness of specific tool features.


Through this research and the development of the JValue project, we argue for the importance of basing both tool and educational design decisions on solid empirical evidence rather than intuition alone. This approach promises to enhance the effectiveness of data engineering tools, learning environments, and consequently, the quality and reliability of data-driven insights across various fields.

To this end, we asked the following research question:

\textit{What are the most common programming mistakes made by beginners in data engineering projects, and what are their consequences?}

Our contributions are as follows:

\begin{enumerate}
    \item The presentation of a systematic review (following Kitchenham et al. \cite{kitchenham2023segress}) to identify and filter pertinent literature,
    \item The quality-assured thematic analysis of the identified-as-relevant literature, and
    \item A resulting classification of common programming mistakes by beginners of data engineering, with implications for tool design and educational strategies.
\end{enumerate}



By identifying and analyzing these mistakes, we provide insights that inform the design of data engineering languages, tools, libraries, and frameworks, as well as educational interventions. We thereby expect to boost the proficiency of novice data engineers. Our analysis includes both general programming mistakes and domain-specific challenges unique to data engineering tasks.

This article is structured as follows. In Section 2 we first review related work. In Section 3 we present the research approach we took and in Section 4 we present the results of the research. In Section 5 we conclude the article and in Section 6 we discuss the limitations of the work presented.

\section{Related Work}
To the best of our knowledge, our work is the first systematic review of common beginner programming mistakes in data engineering.

Prior systematic reviews have examined different aspects of programming education. Qian \& Lehman \cite{10.1145/3077618} reviewed students' misconceptions and difficulties in introductory programming, finding that students struggle with syntactic, conceptual, and strategic knowledge, often due to factors like unfamiliarity with syntax and lack of prior knowledge. Medeiros et al. \cite{medeiros2018systematic} analyzed teaching and learning challenges in introductory programming, highlighting that problem-solving and mathematical ability were the most cited necessary skills, while syntax learning and lack of appropriate teaching methods were common challenges. More recently, Memarian \& Doleck \cite{memarian2024data} focused on data science education specifically, examining pedagogical tools and practices through the lens of technological and pedagogical knowledge quality.

While these previous reviews have broadly covered programming education \cite{10.1145/3077618}, teaching methodologies \cite{medeiros2018systematic}, or data science pedagogy \cite{memarian2024data}, our systematic review addresses a crucial gap by examining both general and data engineering specific programming mistakes, providing a comprehensive view of the challenges faced by beginners in this domain.

\section{Research Approach}
We performed a systematic literature review (SLR) to answer our research question. Our research focuses on student populations as representative of programming beginners, as evidenced by our analyzed literature. The vast majority of empirical studies in this space examine students in introductory programming contexts, with many of these studies explicitly equating students with novice programmers \cite{brown2017novice, brown2014investigating, smith2019error, altadmri201537, jegede2023analysis, hristova2003identifying, junior2019analyzing, denny2012all, ettles2018common, mccall2014meaningful, albrecht2020sometimes, ahmadzadeh2005analysis, kiran2015evaluation, jackson2005identifying, mccall2019new, miedema2022identifying}.

We conducted our systematic review following established guidelines for systematic literature reviews \cite{kitchenham2004procedures,booth2021systematic,webster2002analyzing}, while structuring our reporting according to the SEGRESS guidelines \cite{kitchenham2023segress}. For the analysis of the identified literature, we employed Braun \& Clarke's \cite{braun2012thematic} thematic analysis method to derive a theory from the data. The resulting theory takes the form of a classification of programming mistakes made by beginners.

\begin{figure}[h]
\centering
\includegraphics[width=0.8\columnwidth]{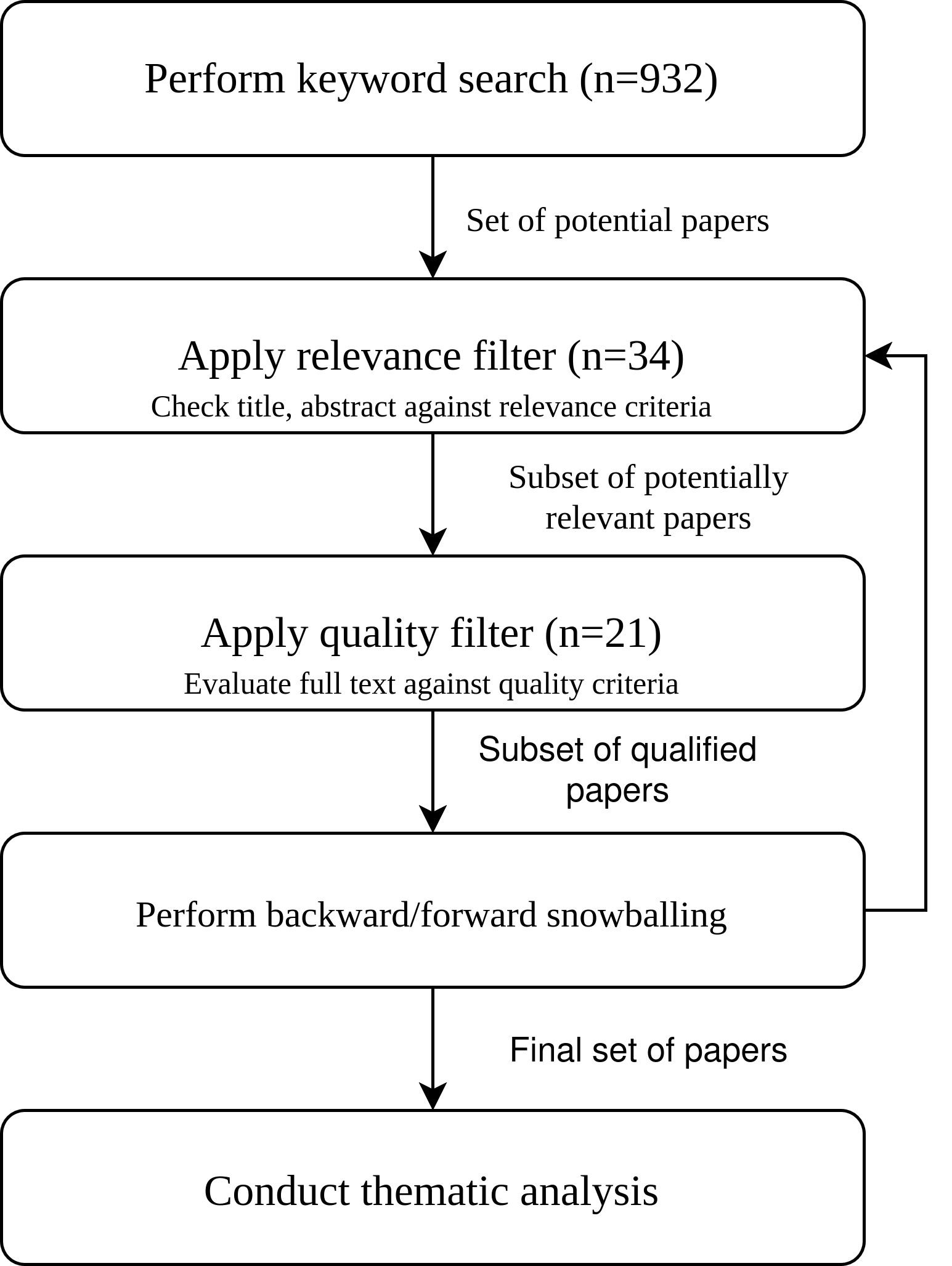}
\caption{Visualization of data extraction and synthesis}
\label{researchApproach}
\end{figure}

Figure \ref{researchApproach} illustrates the iterative research process. The process begins with performing a keyword search, which results in an initial set of papers. From this list of potential papers, we apply a relevance filter, checking titles and abstracts against relevance criteria. This produces a subset of potentially relevant papers.

Next, we applied a quality filter, evaluating the full text against quality criteria, resulting in a subset of qualified papers. These qualified papers then undergo thematic analysis. Throughout the process, we perform backward and forward snowballing to identify additional relevant articles \cite{wohlin2014guidelines}.

\subsection{Literature Search}

Our search approach utilized two major electronic databases: IEEE Xplore and Scopus. We developed two search strings to capture both data engineering-specific and general programming errors that could be relevant to data engineering:

\begin{enumerate}
    \item ("data engineering" OR "data science") AND ("mistake" OR "fault" OR "error" OR "misconception" OR "bug") AND ("beginner" OR "novice" OR "student")
    \item "programming" AND ("mistake" OR "error") AND ("beginner" OR "novice" OR "student") AND "data"
\end{enumerate}

The first search string, focusing specifically on data engineering and data science mistakes, yielded 98 results in Scopus and 107 in IEEE Xplore. The second search string, which captured broader programming mistakes in data-related contexts, returned 458 results in Scopus and 269 in IEEE Xplore.



\subsection{Initial Selection Criteria}
Before applying our detailed relevance and quality filters, we established the following practical criteria for study selection:

\begin{itemize}
\item Language: We limited our review to English-language publications due to the researchers' language capabilities and to ensure consistent interpretation of technical terminology.

\item Accessibility: All relevant papers were accessible through institutional subscriptions or were openly available, requiring no direct author contact.

\item Publication Type: We included only peer-reviewed publications and full conference papers to ensure a baseline of scientific rigor, excluding non-peer-reviewed publications and brief conference abstracts.
\end{itemize}


\subsection{Inclusion and Exclusion Criteria}
We adopted a two-phase approach to assess the relevance and quality of the identified studies. 

\subsubsection{Relevance Filter}
The relevance filter was applied to ensure that the selected studies aligned with the core focus of our research. The criteria for this phase were as follows:
Inclusion Criteria:

\begin{itemize}
    \item Topic Relevance: Studies were included if they met one of the following criteria:
        \begin{itemize}
            \item Focused directly on programming mistakes in data science or data engineering contexts
            \item Provided comprehensive analysis of novice programming mistakes in languages commonly used in data engineering (Python, R, SQL, Java)
            \item Offered insights into fundamental programming concepts and error patterns that impact data manipulation, analysis, and processing tasks
        \end{itemize}
    
    \item Target Audience: Research should address novice programmers.
    
    \item Research Type: Empirical studies, case studies, theoretical papers, reviews, and meta-analyses contributing to understanding or defining best practices in error prevention or management in programming were included.
    
    \item Educational Context: Papers discussing lessons learned from teaching data engineering classes were considered relevant, if they offered insights into the programming mistakes.
\end{itemize}

    
    

\subsubsection{Quality Filter}
Studies that passed the relevance filter were then subjected to a rigorous quality assessment. The criteria for this phase were as follows:

Inclusion Criteria:
\begin{itemize}
    \item Methodological Rigor: Studies must demonstrate clear research design and systematic data collection methods. For quantitative studies, this included appropriate sample sizes and statistical analyses. For qualitative studies, this meant well-documented methodological approaches such as grounded theory, thematic analysis, or case study protocols with clear data collection and analysis procedures.
    
    \item Publication Quality: Peer-reviewed publications in journals or conference proceedings in the field of data engineering or computer science were included.
    
    \item Comprehensive Analysis: Studies providing in-depth analysis of programming mistakes, including causes, consequences, and potential solutions, along with detailed discussions of findings, practical implications, and future research directions were favored.
\end{itemize}

Exclusion Criteria:
\begin{itemize}
    \item Lack of Empirical Data: Studies relying solely on anecdotal evidence or opinion pieces without empirical support were excluded.
    
    \item Methodological Weaknesses: Research with unclear designs, inadequate sample sizes, or insufficient detail on data analysis methods was excluded. Studies that did not address validity and reliability issues were also omitted.
    
\end{itemize}

To provide a nuanced evaluation of each paper's relevance and quality, a scoring system was implemented. Each criterion was marked as "yes" (2 points), "partial" (1 point), or "no" (0 points). The final score for each paper was calculated as the square root of the sum of squared values for each criterion.
This two-phase approach ensured a comprehensive and rigorous selection process, aiming to provide a holistic view of programming mistakes in data engineering.

\subsection{Data Extraction and Analysis}
Selected papers were accessed through various online academic platforms using university credentials. The qualitative data analysis tool QDAcity\footnote{See \url{https://qdacity.com}}
 was used for detailed data analysis.

We employed Braun \& Clarke's \cite{braun2012thematic} thematic analysis approach to systematically identify, analyze, and report patterns (themes) within the data. This method involves six phases:

\begin{enumerate}
    \item Familiarization with the data: We thoroughly read and re-read the selected papers, making initial notes on potential codes and themes.
    
    \item Generating initial codes: We systematically coded interesting features across the entire dataset, collating data relevant to each code.
    
    \item Searching for themes: We collated codes into potential themes, gathering all data relevant to each potential theme.
    
    \item Reviewing themes: We checked if the themes work in relation to the coded extracts and the entire dataset, generating a thematic 'map' of the analysis.
    
    \item Defining and naming themes: We conducted ongoing analysis to refine the specifics of each theme, generating clear definitions and names for each theme.

    \item Producing the report: We selected compelling extract examples, conducted final analysis of selected extracts, related the analysis back to the research question and literature, and produced a scholarly report of the analysis.
\end{enumerate}

The analysis involved iterative coding of the papers to identify and categorize different types of programming mistakes. This dynamic process allowed for continuous refinement and redefinition of categories as new patterns emerged from the data. We employed an inductive approach in our coding process \cite{corbin2014basics}. This method allowed themes to emerge organically from the data itself, without preconceived notions or existing frameworks. As we progressed through the analysis, we consistently revisited and refined our categorizations to ensure they accurately reflected the patterns observed in the collected data.

To ensure the reliability and validity of our analysis, we employed several strategies:

\begin{itemize}   
    \item Constant comparison: We continuously compared new data with existing codes and themes, refining our analysis throughout the process \cite{corbin2014basics}.
    
    \item Negative case analysis: We actively searched for and discussed cases that did not fit within the emerging patterns to ensure a comprehensive analysis \cite{lincoln1985naturalistic}.
\end{itemize}

Throughout the analysis process, both authors held regular review meetings to discuss emerging themes, validate coding decisions, and ensure consistent interpretation of the data.

Our use of thematic analysis aligns with best practices in qualitative research in software engineering \cite{kitchenham2004procedures, corbin2014basics}, allowing for a rigorous and systematic exploration of the literature. The resulting themes form the basis of our classification of programming mistakes, providing a structured framework for understanding and addressing common challenges faced by novice data engineers.

\begin{figure}[h]
\centering
\includegraphics[width=0.8\columnwidth]{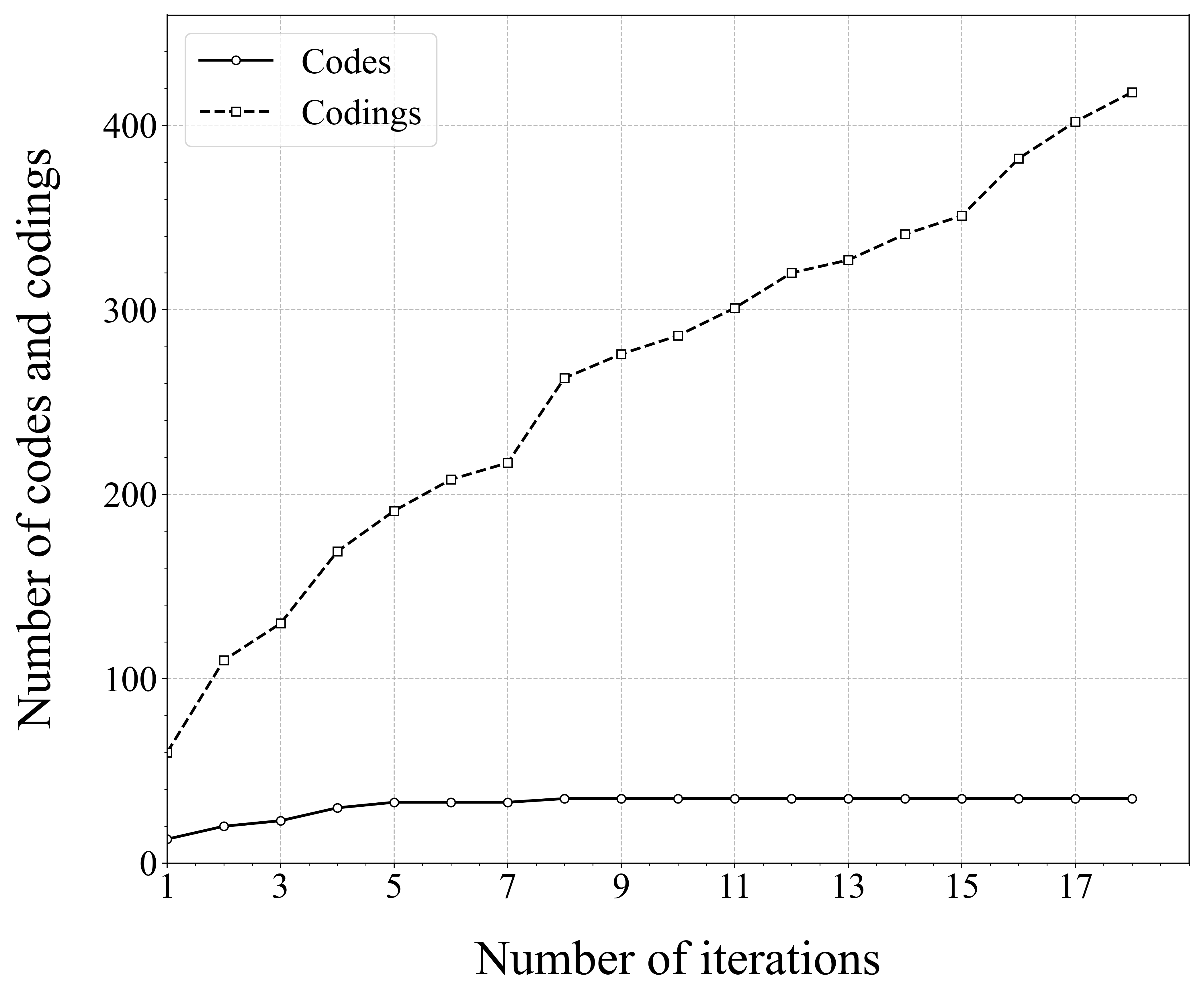}
\caption{Codes and codings over time, showing theoretical saturation \cite{bowen2008naturalistic}}
\label{fig:saturation}
\end{figure}

Figure \ref{fig:saturation} illustrates the progression of our coding process over time. The graph shows the cumulative number of unique codes (solid line) and total codings (dotted line) as we analyzed each paper. The declining and eventual lack of growth of codes over time indicates theoretical saturation, suggesting that we reached a point where additional data analysis was not yielding new insights or categories. This saturation provides confidence in the comprehensiveness of our thematic analysis and the resulting classification of programming mistakes.


\section{Results}

This literature review analyzed 21 publications spanning from 2003 to 2024, comprising 6 journal articles, 14 conference proceedings papers, and 1 symposium article (see \ref{appendix:literature}). Given the limited literature specifically addressing programming mistakes in data engineering, our review encompasses both domain-specific studies and broader research on beginner programming mistakes that are relevant to data science contexts.

\begin{figure}[h]
\centering
\includegraphics[width=0.9\columnwidth]{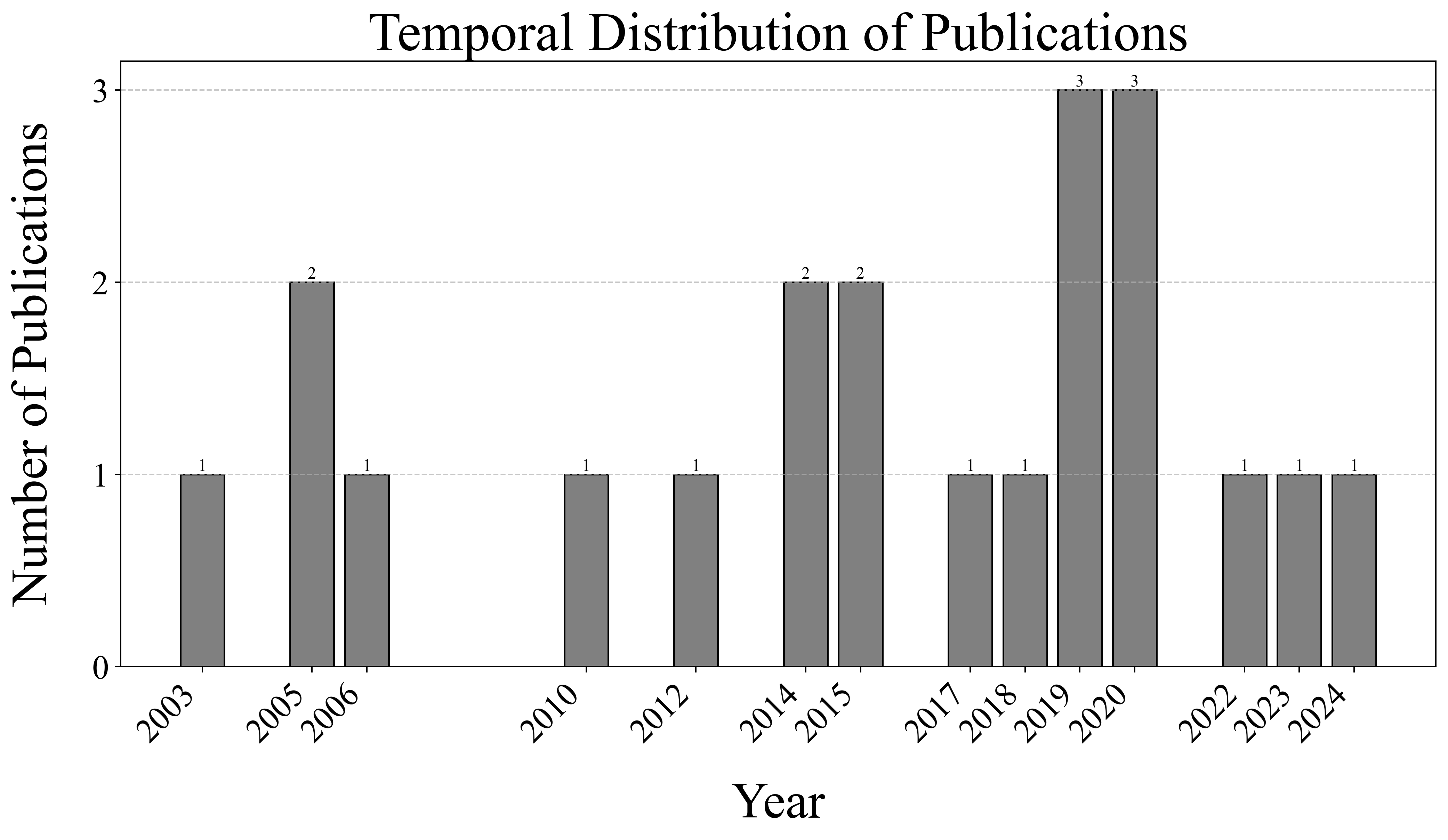}
\caption{Distribution of Publications by Year}
\label{fig:year_dist}
\end{figure}

The temporal distribution of these publications, shown in Figure \ref{fig:year_dist}, reveals the evolution of research interest in this area.

The broader literature covers a diverse range of programming languages illustrated in figure \ref{fig:languages}:

\begin{figure}[h]
\centering
\includegraphics[width=0.6\columnwidth]{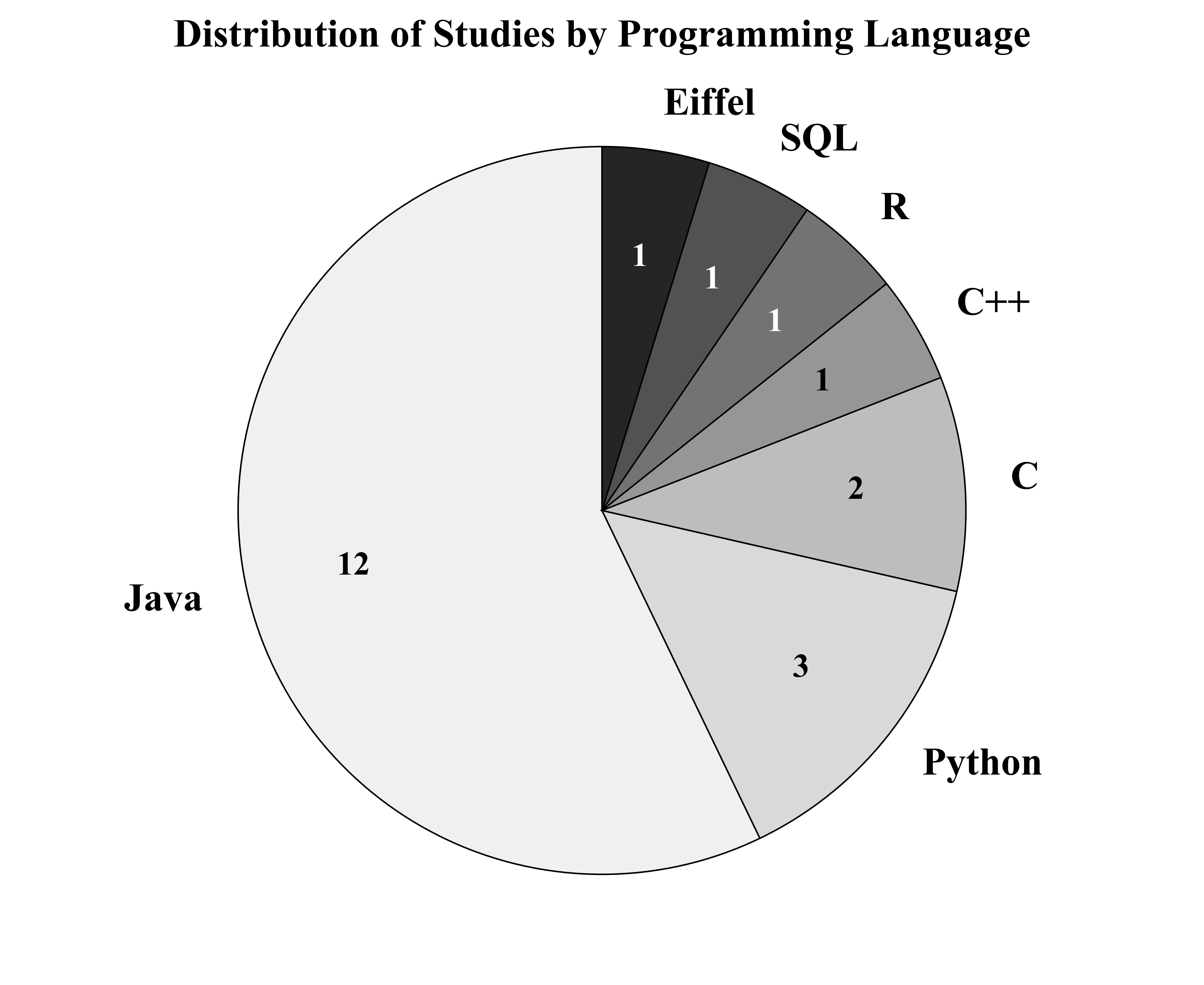}
\caption{Distribution of Studies by Programming Language}
\label{fig:languages}
\end{figure}

Our analysis shows Java as the predominant programming language in the studied papers, followed by Python. This aligns with recent industry research by 365 Data Science \cite{magnet2024} which found that SQL, Python, and Java remain the most crucial programming languages in data engineering, with SQL appearing in 79.4\% of job postings, Python in 73.7\%, and Java in 22.6\% of listings. Java's strong presence in our academic literature likely reflects its historical importance in building scalable data processing systems and its widespread use in enterprise data engineering tools like Apache Hadoop and Apache Spark.

The research methodologies employed in these studies reveal a diverse range of approaches, with 13 quantitative studies and 8 mixed-methods studies. The quantitative studies can be further divided into three main subcategories:

\begin{itemize}
\item Post-hoc data analysis: 6 studies analyzed programming logs to identify error patterns and frequencies \cite{brown2014investigating, brown2017novice, smith2019error, denny2012all, ahmadzadeh2005analysis, yarygina2020learning}.
\item Large-scale analysis: 3 studies conducted extensive error analyses in controlled environments \cite{altadmri201537, hristova2003identifying, vee2006empirical}.
\item Targeted analysis of specific programming tasks: 4 studies focused on detailed evaluations of specific assignments or programming patterns \cite{kaczorowska2020analysis, junior2019analyzing, jackson2005identifying, ettles2018common}.
\end{itemize}

These quantitative studies primarily utilized numerical data and statistical methods to identify trends and draw conclusions. The remaining 8 mixed-methods studies \cite{singh2024investigating, jegede2023analysis, mccall2014meaningful, mccall2019new, kaczmarczyk2010identifying, kiran2015evaluation, miedema2022identifying, albrecht2020sometimes} combined quantitative techniques with qualitative methods such as thematic analysis, interviews, and think-aloud protocols. This methodological diversity enabled us to identify both general programming mistakes that persist across contexts and those specific to data engineering tasks.


The following sections provide a detailed analysis of the common mistakes made by beginners in data engineering, as summarized in Table \ref{tab:results}. These sections are organized according to the main themes identified in our research. Each section explains a specific mistake type within these categories, providing definitions, examples, and discussions to offer a comprehensive understanding of the challenges faced by novice data engineers. A detailed mapping between each identified mistake type and the supporting literature can be found in Appendix \ref{appendix:mapping}.

\subsection{\textbf{Domain-Specific Mistakes (1)}}
The Domain-Specific Mistakes are about programming mistakes specific to data engineering and data science.

\textbf{(1.1) Dataset Misunderstandings \label{mistake:dataset-misunderstandings}} \\
Definition: Mistakes arising from misunderstanding the structure, schema, or specific attributes of the dataset. \\
Examples: \\
- \textit{Incorrect Attribute Selection}: In one study \cite{singh2024investigating}, a common mistake was selecting the wrong dataset attribute. Students often used the attribute 'PDAT' instead of 'P\_NUMVRC', misinterpreting 'PDAT' to indicate the number of varicella doses received rather than adequate provider data. \\
- \textit{Incorrect Value Usage}: Another frequent mistake involved using incorrect dataset values. For example, in the same study \cite{singh2024investigating}, students used 'male' and 'female' instead of numerical codes (1 and 2) when filtering rows by gender. \\

\textbf{(1.2) Faulty Data Analysis Logic \label{mistake:faulty-data-analysis-logic}} \\
Definition: Mistakes in implementing data analysis algorithms or logical data processing. \\
Examples: \\
- \textit{Incorrect Use of DataFrame Functions}: In one instance \cite{singh2024investigating}, students incorrectly used the where() function from pandas to select rows, which "does not filter out the rows where the condition is not satisfied, as where() is used for replacing values where the condition is False to some specific value (specified in the call through a parameter named other). Consequently, in the subsequent lines of code, len(df) returns the length of the full DataFrame rather than the number of rows where the condition is satisfied." \\
- \textit{Incorrect DataFrame Operations}: Another example involved "dividing a DataFrame slice by another slice rather than dividing their lengths to compute a given ratio." \cite{singh2024investigating} \\

\textbf{(1.3) Incorrect Data Handling \label{mistake:incorrect-data-handling}} \\
Definition: Errors in manipulating or accessing data, leading to incorrect data processing. \\
Examples: \\
- \textit{Incorrect Conditions}: An example from \cite{singh2024investigating} involved using \texttt{df['P\_NUMVRC']!=0} instead of \texttt{df['P\_NUMVRC']>0} to select rows indicating children who received at least one varicella dose. This condition fails because it does not correctly account for missing values, which are counted in addition to the number of rows where 'P\_NUMVRC' is greater than 0. \\
- \textit{Not Accounting for NaN Values}: Another common mistake was replacing all missing values with 0, which led to incorrect calculations. For instance, when computing the correlation between the 'P\_NUMVRC' and 'HAD\_CPOX' columns, a student replaced all missing values in both columns with 0, leading to incorrect computation of the correlation as missing values from the 'P\_NUMVRC' column were incorrectly counted towards children receiving zero doses. \cite{singh2024investigating} \\
- \textit{Misunderstanding Function Outputs}: Students also made mistakes in using the groupby() function incorrectly, where they misunderstood the output format of the object returned by groupby and performed incorrect operations in subsequent lines of code. \cite{singh2024investigating} \\

\textbf{(1.4) Misunderstanding Data Types \label{mistake:misunderstanding-data-types}} \\
Definition: Incorrect handling or interpretation of different data types. \\
Examples: \\
- \textit{Format Errors}: Students often made errors related to data formats, such as trying to substitute categorical data into a histogram, which is not appropriate. An example error message observed was: "Error in mutate-impl(.data, dots) : Evaluation error: non-numeric argument to binary operator". \cite{yarygina2020learning}

\subsection{\textbf{Strategic Errors (2)}}

\textbf{(2.1) Suboptimal Coding \label{mistake:suboptimal-coding}} \\
Definition: Inefficient use of coding constructs, leading to suboptimal performance, especially with large datasets. This includes not using appropriate vectorized functions for data manipulation. \\
Examples: \\
- \textit{Inefficient Loop Usage}: A prevalent issue identified in multiple papers (\cite{singh2024investigating}, \cite{albrecht2020sometimes}) is the use of for-loops to iterate over DataFrame rows, instead of using vectorized operations. For instance, students used for-loops to count rows satisfying a condition, which is significantly slower than applying a Boolean mask. \\
- \textit{Complex Solutions}: Some students submitted unnecessarily complex solutions for problems that could be solved with simpler, more efficient code. This was observed in \cite{albrecht2020sometimes}, where students wrote complex, multi-line solutions instead of concise, optimal code. \\

\onecolumn

\begin{longtable}{|>{\centering\arraybackslash}p{3.6cm}|>{\centering\arraybackslash}p{4.5cm}|>{\centering\arraybackslash}p{7cm}|>{\centering\arraybackslash}p{1.8cm}|}
    \caption{Summary of Common Beginner Programming Mistakes in Data Engineering.}
    \label{tab:results} \\
    \hline
    \textbf{Top-level Code} & \textbf{Mistake Type} & \textbf{Definition} & \textbf{Occurrences} \\
    \hline
    \endfirsthead



    \hline
    \endlastfoot

    \textbf{Domain-Specific Mistakes (1)} & (1.1) Dataset Misunderstandings \label{table-mistake:dataset-misunderstandings} & Mistakes arising from misunderstanding the dataset, its schema, or associated data guides. & 1 \\
     & (1.2) Faulty Data Analysis Logic \label{table-mistake:faulty-data-analysis-logic} & Mistakes in implementing data analysis algorithms. & 2 \\
     & (1.3) Incorrect Data Handling \label{table-mistake:incorrect-data-handling} & Mistakes in manipulating data. & 2 \\
     & (1.4) Misunderstanding Data Types \label{table-mistake:misunderstanding-data-types} & Incorrect handling of different data types. & 2 \\
    \hline
    \textbf{Strategic Errors (2)} & (2.1) Suboptimal Coding \label{table-mistake:suboptimal-coding} & Inefficient use of coding constructs. & 3 \\
     & (2.2) Wrong Algorithm \label{table-mistake:wrong-algorithm} & Choosing the incorrect algorithm for the task. & 5 \\
    \hline
    \textbf{Misinterpretation Errors (3)} & (3.1) Incorrect Programming Assumptions \label{table-mistake:incorrect-programming-assumptions} & Making wrong assumptions about how programming constructs work. & 10 \\
     & (3.2) Task Misunderstanding \label{table-mistake:task-misunderstanding} & Misinterpreting the problem statement or requirements. & 3 \\
    \hline
    \textbf{Environment and Language Misconceptions (4)} & (4.1) Incorrect Language Understanding \label{table-mistake:incorrect-language-understanding} & Misunderstandings about the overall functionality, rules, and behaviors of the programming language, as well as the development environment in which novices work. & 18 \\
     & (4.2) Library Misuse \label{table-mistake:library-misuse} & Incorrect use of libraries or failure to import necessary libraries. & 4 \\
     & (4.3) Path and I/O Errors \label{table-mistake:path-io-errors} & Incorrectly specifying file paths or managing file I/O. & 4 \\
    \hline
    \textbf{Logical Errors (5)} & (5.1) Incorrect Loop Conditions \label{table-mistake:incorrect-loop-conditions} & Errors in loop conditions, leading to infinite loops or off-by-one errors. & 2 \\
     & (5.2) Faulty Conditional Logic \label{table-mistake:faulty-conditional-logic} & Errors in if-else statements that cause incorrect branching. & 3 \\
     & (5.3) Off-by-One / Out-of-Bounds Errors \label{table-mistake:off-by-one-out-of-bounds-errors} & Misplacing indices in loops or arrays. & 6 \\
    \hline
    \textbf{Semantic Errors (6)} & (6.1) Incorrect Function Usage \label{table-mistake:incorrect-function-usage} & Using functions incorrectly. & 12 \\
     & (6.2) Type Mismatches \label{table-mistake:type-mismatches} & Assigning or passing incorrect data types to variables or functions. & 15 \\
     & (6.3) Variable Scope Issues \label{table-mistake:variable-scope-issues} & Using variables outside their defined scope. & 5 \\
    \hline
    \textbf{Syntax Errors (7)} & (7.1) Incorrect Loop \& if-else Syntax \label{table-mistake:incorrect-loop-if-else-syntax} & Errors in the syntax of loops or if else statements. & 5 \\
     & (7.2) Incorrect Operators \label{table-mistake:incorrect-operators} & Using operators incorrectly. & 12 \\
     & (7.3) Invalid Keywords Usage \label{table-mistake:invalid-keywords-usage} & Misusing language keywords. & 3 \\
     & (7.4) Missing Semicolons \label{table-mistake:missing-semicolons} & Omitting semicolons where required. & 7 \\
     & (7.5) Unbalanced Delimiters \label{table-mistake:unbalanced-delimiters} & Unbalanced brackets, parentheses, or braces. & 11 \\
     & (7.6) Variable not declared or initialized \label{table-mistake:variable-not-declared-or-initialized} & Using variables that are not declared or initialized. & 9 \\
    \hline
    \textbf{Sloppiness Errors (8)} & (8.1) Typographical Errors \label{table-mistake:typographical-errors} & Simple typos in code. & 7 \\
    \hline
    \textbf{Memory Management Mistakes (9)} & (9.1) Memory Language Misconceptions \label{table-mistake:memory-language-misconceptions} & Misunderstanding memory management in programming languages. & 3 \\
     & (9.2) Memory Leaks \label{table-mistake:memory-leaks} & Failing to deallocate memory. & 2 \\
    \hline
\end{longtable}


\begin{multicols}{2}

\textbf{(2.2) Wrong Approach \label{mistake:wrong-algorithm}} \\
Definition: Using an inappropriate or fundamentally flawed algorithm to solve a problem, which often leads to incorrect results or inefficient solutions. \\
Examples: \\
- \textit{Incorrect Ratio Computation}: An example from \cite{singh2024investigating} involved a student computing a ratio by dividing one DataFrame slice by another, rather than dividing their lengths. This fundamental misunderstanding of the problem led to incorrect results. \\
- \textit{Incorrect Maximum Value Calculation}: In another example from \cite{ettles2018common}, a student's algorithm failed to correctly handle edge cases in determining the maximum value in an array, leading to incorrect outputs when all values were negative. \\

\subsection{\textbf{Misinterpretation Errors (3)}}

\textbf{(3.1) Incorrect Programming Assumptions \label{mistake:incorrect-programming-assumptions}} \\
Definition: Misconceptions about how specific programming constructs or functions work. \\
Examples: \\
- \textit{Bitwise Operators Misuse}: In \cite{singh2024investigating}, students incorrectly used the bitwise AND operator on DataFrame objects, which is not valid. The correct approach involves combining Boolean masks with appropriate precedence rules. \\
- \textit{Incorrect Function Assumptions}: Another example from \cite{albrecht2020sometimes} involved students misunderstanding how functions like \texttt{getchar()} and \texttt{scanf()} handle input buffers, leading to persistent errors in their programs. \\

\textbf{(3.2) Task Misunderstanding \label{mistake:task-misunderstanding}} \\
Definition: Misinterpreting the problem statement, leading to incorrect implementation. \\
Examples: \\
- \textit{Ratio Calculation Errors}: In \cite{singh2024investigating}, students misunderstood the problem requirements, calculating ratios incorrectly by including all vaccinated children instead of only those who did not contract chickenpox. \\
- \textit{Incorrect Output Format}: Papers \cite{singh2024investigating}, \cite{junior2019analyzing}, \cite{albrecht2020sometimes}, \cite{kaczmarczyk2010identifying}, \cite{ettles2018common} noted that students often misunderstood the required output format, leading to errors like returning rounded ratios instead of exact values or incorrect dictionary key ordering. \\

\subsection{\textbf{Environment and Language Misconceptions (4)}}

\textbf{(4.1) Incorrect Language Understanding \label{mistake:incorrect-language-understanding}} \\
Definition: Making incorrect assumptions about how the used programming language works, leading to fundamental errors. This is a major issue mentioned across many studies and examples. \\
Examples: \\
- \textit{Misunderstanding Control Flow}: A common error is control flow reaching the end of a non-void method without returning a value. For example, in \cite{brown2017novice}, a method was defined as follows:
\begin{verbatim}
public int foo(int x) {
    if (x < 0) return 0;
    x += 1;
}
\end{verbatim}
This leads to a compilation error because the method does not return a value in all code paths. \\
- \textit{Incorrect Method Invocation}: In \cite{brown2017novice}, students often attempted to invoke non-static methods as if they were static. \\
- \textit{Parameter Types in Method Calls}: Including parameter types when invoking a method is another error noted in \cite{brown2017novice}. \\
- \textit{Misunderstanding Scope Rules}: In Python, misunderstanding scope rules can lead to errors like referencing a variable before assignment. 

\textbf{(4.2) Library Misuse \label{mistake:library-misuse}} \\
Definition: Incorrect use of libraries, such as failing to import them correctly or misunderstanding their usage. \\
Examples: \\
- \textit{Missing Imports}: As seen in \cite{singh2024investigating}, students often forgot to import necessary libraries, which led to runtime errors. An example error is failing to import `pandas` but still attempting to use `pd.DataFrame`.  \\
- \textit{Improper Use of Libraries}: Another example is not defining commonly used variables or aliases (like `df` for DataFrame), causing confusion and errors in subsequent code cells. \\

\textbf{(4.3) Path and I/O Errors \label{mistake:path-io-errors}} \\
Definition: Errors related to incorrectly specifying file paths or managing input/output operations. \\
Examples: \\
- \textit{Hardcoded Paths}: Students often hardcoded absolute paths that were inaccessible on other machines, such as:
\begin{verbatim}
import pandas as pd
df = pd.read_csv('/path/to/data.csv')
\end{verbatim}
- \textit{Misunderstanding Input Operations}: Spohrer and Soloway \cite{spohrer1986novice} found that novices often misunderstood how input operations handle whitespace. For instance, some students failed to understand that whitespace in the input data is a character that can be read, not just a separator that is automatically ignored. This high-frequency bug was caused by a misunderstanding of how the READLN statement works in the context of characters. Some novices thought that READLN ignores all whitespace when parsing an input line and assigning values to a sequence of variables, possibly because that's how READLN works on a sequence of numeric inputs.



\subsection{\textbf{Logical Errors (5)}} \label{mistake:logical-errors}
Logical errors are mistakes in the code's logic that lead to incorrect behavior. Common examples include incorrect loop conditions, which can result in infinite loops or off-by-one errors, and faulty conditional logic in if-else statements, leading to incorrect branching. Notably, off-by-one or out-of-bounds errors, where indices in loops or arrays are misplaced, occur frequently and can cause significant issues by accessing elements outside their intended range.

\subsection{\textbf{Semantic Errors (6)}} \label{mistake:semantic-errors}
Semantic errors in data engineering arise from the misuse of functions, operators, or data types, leading to incorrect program behavior. A significant issue in this category is the incorrect usage of functions, particularly regarding the number of arguments and their types. This mistake is especially prevalent among novices and can lead to numerous errors. Type mismatches, which involve assigning or passing incorrect data types to variables or functions, are almost universally encountered in the literature and represent a major source of misunderstanding for beginners learning a programming language. Additionally, variable scope issues, where variables are used outside their defined scope, can result in unexpected behavior.

\subsection{\textbf{Syntax Errors (7)}} \label{mistake:syntax-errors}
Syntax errors in data engineering are mistakes that violate the grammatical rules of the programming language, preventing code from compiling or running. These errors are ubiquitous and occur frequently, but are typically quick and easy to fix. They include issues such as missing semicolons, incorrect operators (e.g., confusing assignment with comparison), errors in loop and if-else syntax, invalid keyword usage, and undeclared or uninitialized variables. Among these, unbalanced delimiters - such as mismatched parentheses, curly braces, or square brackets - stand out as the most significant and common error. Despite their prevalence, the relative ease of identifying and correcting syntax errors makes them less problematic in the long run compared to logical or semantic errors. However, their frequency can still significantly impact productivity and learning curves for novice data engineers, highlighting the importance of robust syntax checking tools and clear error messages in development environments.

\subsection{\textbf{Sloppiness/Typographical Errors (8)}} \label{mistake:typographical-errors}
Sloppiness or typographical errors are mistakes that occur due to carelessness or simple typing errors in the code. These errors can afflict programmers of all skill levels, from novices to experts. While often easy to fix once detected, they can create a cascade of problems if left unnoticed. Typos in variable names, function calls, or numerical values can lead to unexpected behavior, logical errors, or even syntax errors. The insidious nature of these mistakes lies in their potential to introduce subtle bugs that may not be immediately apparent, potentially causing significant issues in data processing or analysis down the line. However, with proper attention to detail and the use of code review practices or linting tools, these errors can usually be caught and corrected quickly, minimizing their impact on the overall data engineering process.

\subsection{\textbf{Memory Management Errors (9)}} \label{mistake:memory-anagement-errors}
Memory management errors occur when programmers mishandle memory allocation and deallocation, which, though less frequently discussed, can have significant impacts. Two common mistakes include \textit{memory language misconceptions} and \textit{memory leaks}.

Memory misconceptions arise when students assume that declared objects are automatically allocated memory without instantiation \cite{kaczmarczyk2010identifying} or attempt to access memory via unallocated pointers \cite{albrecht2020sometimes}. These misunderstandings can lead to hard-to-detect bugs.

Memory leaks occur when students fail to deallocate allocated memory, leading to resource wastage and potential system slowdowns \cite{albrecht2020sometimes}. This mistake can have serious performance consequences, especially in long-running applications.




\section{Discussion}
The findings of this study reveal a complex landscape of programming mistakes encountered by novices. While some mistakes are common across various programming domains, others are particularly unique to data engineering tasks.

\subsection{Discussion of programming mistakes}
General programming mistakes, such as syntax mistakes involving missing semicolons or unbalanced delimiters (\ref{mistake:syntax-errors}), are omnipresent among novice programmers regardless of their specific field. Similarly, logical errors related to conditional statements, loops, or off-by-one errors are frequent occurrences (\ref{mistake:logical-errors}). These types of mistakes, while common, often represent opportunities for targeted instruction and early intervention in educational settings, where students could benefit from exercises and real-time feedback that reinforce proper syntax and logical reasoning skills.

More significant and persistent challenges emerge in areas specific to data engineering. A primary concern is the misunderstanding or mishandling of datasets (\ref{mistake:dataset-misunderstandings}). This issue often arises due to a lack of familiarity with data structures, insufficient domain knowledge, or inadequate understanding of data manipulation techniques. In educational contexts, addressing these issues through curriculum adjustments, such as including more hands-on training with diverse datasets and focused instruction on data processing techniques, could help students build a deeper understanding. By incorporating problem-based learning or interactive simulations into teaching, educators can provide practical exposure that reduces the likelihood of these errors.

Another critical area of difficulty lies in algorithm selection and implementation. Novices frequently struggle to identify the most suitable algorithms for specific data engineering tasks and may face challenges in correctly implementing these algorithms (\ref{mistake:wrong-algorithm}). This issue is often compounded by inefficient coding practices, such as failing to utilize vectorized functions in favor of less efficient manual loops (\ref{mistake:suboptimal-coding}). Educators could address these gaps by designing curricula that focus on algorithmic thinking and computational efficiency, using real-world examples to show the impact of suboptimal practices. Educational tools that visualize the performance differences between different coding strategies can also help students better understand these concepts.

Data type handling and file path management present particularly relevant challenges in the data engineering context. Given the diverse nature of data sources and formats, students must develop a strong understanding of data types (\ref{mistake:misunderstanding-data-types}) and file handling (\ref{mistake:path-io-errors}). Educational interventions that provide students with a variety of data-handling scenarios could help mitigate these issues. For example, structured labs or assignments that focus on reading, writing, and managing data from different sources can reinforce these critical skills, preparing students to handle real-world data challenges.

Many mistakes also simply stem from a basic lack of understanding of the programming language and environment. This deficit manifests in various ways, from misunderstanding basic programming constructs to incorrectly assuming how specific functions operate. In educational settings, these issues could be addressed by incorporating detailed instruction on language-specific quirks and encouraging exploratory learning, where students test their assumptions in sandbox environments. Educators should also emphasize the importance of libraries and memory management, which are critical for efficient data processing, through hands-on labs and debugging exercises.

Interestingly, this study highlights that novices frequently struggle with interpreting problem statements correctly (\ref{mistake:task-misunderstanding}). This finding underscores the importance of developing not only technical skills but also analytical and comprehension abilities. Educators could address this challenge by incorporating assignments that require students to break down and restate complex problems, thus ensuring a solid understanding of the task at hand. This approach could help students learn how to approach data engineering tasks methodically, reducing the risk of misinterpretation.

\subsection{Implications for educators and industry}

These findings have several practical implications for both tool development and educational practices in data engineering. Integrated Development Environments (IDEs) can be enhanced with features that specifically target common novice errors in data engineering. For example, advanced debugging tools, real-time feedback on data manipulation, and suggestions for efficient coding practices could potentially reduce the occurrence of these mistakes. Similarly, educational platforms could integrate these features to give students immediate feedback during coding exercises, reinforcing correct practices while they learn.

In the industry, companies should consider incorporating these findings into their onboarding programs for new data engineers. This could include workshops on common pitfalls and best practices in data engineering, ensuring that novices are well-prepared to handle the challenges they will face. Likewise, educational institutions should incorporate this knowledge into their curricula, aligning their teaching with real-world challenges to better prepare students for professional environments. Establishing strong mentorship programs, encouraging peer code reviews, and utilizing interactive learning platforms can help novices in both academia and industry environments learn from experienced data engineers, reducing the frequency of common mistakes.

By addressing these common mistakes through targeted educational interventions, tool features, and supportive industry practices, we can significantly enhance the proficiency and effectiveness of novice data engineers. This, in turn, will lead to more reliable and accurate data processing, ultimately contributing to better data-driven decision-making across various fields.

\section{Limitations}

We discuss the limitations of our work using Guba \& Lincoln’s \cite{lincoln1982establishing} quality criteria for qualitative research. The four quality criteria for qualitative research (credibility, transferability, dependability, and confirmability) mirror the traditional four quality criteria for quantitative research (internal validity, external validity, reliability, and objectivity).

\subsection {Credibility}

The credibility of findings in qualitative research (internal validity in quantitative research) rests on the connection (ideally isomorphism) between the studied phenomena and the empirical data gathered about the phenomena. Data quality assurances like prolonged engagement and persistent observation support credibility and are fulfilled in our work through the accuracy of our search queries and selection filters: We captured all work (to the extent that search engines could find it) and hence identified all relevant data we needed for our analysis, similarly to how prolonged engagement and persistent observation would have afforded it to an investigator of a primary study (rather than a secondary study like ours).

A strength of a systematic review is the built-in triangulation afforded by the different data sources (articles) utilized. Each article provides both investigator triangulation (different people’s work) and data triangulation (different underlying data sets for the findings) to stabilize the findings. Even if any article's particular findings were off target, the rest would reign them in. The theoretical saturation our data analysis achieved shows that our work, based on this broad and deep primary data, would not have found much to add if we had continued, and thereby shows completion of our analysis.

\subsection{Transferability}

The transferability of findings in qualitative research (external validity in quantitative research) is about changing contexts and still being able to apply the findings. Transferability is less important for us; we believe the goal of improving data engineering is wholly sufficient for our work. Still, the main quality criterion of a valid context transfer, called thick description, is built-in into the empirical data (the articles) our research is built on. The breadth and depth of data and interpretation in our primary materials is ensured by the quality of the research publications we identified and built the systematic review from. The combination of having exhausted the search space (available articles) together with having finished the possible interpretation (theoretical saturation) shows that our data was as thick as it could get.
It's important to note that most of our findings are derived from studies of beginner student courses. The transferability of these findings to beginners or novices in professional data engineering contexts may be limited. The challenges faced by students in controlled academic environments may differ from those encountered by professionals learning on the job. Additionally, our findings may not fully represent all aspects of data engineering, as the field encompasses a broad range of skills beyond programming, including data modeling, Extract Transform Load (ETL) processes, and data pipeline management.

\subsection{Dependability}

The dependability of findings in qualitative research (reliability in quantitative research) is the stability of findings after all random variation has been removed. Here, a systematic review really shines to the extent that the search query and selection filters allowed the investigators to identify all relevant work. The replication of the same search query and article selection at a different point in time would only be different for the primary data to the extent that new work would have been found (as is possible at a future point in time). 

\subsection{Confirmability}

The confirmability of findings in qualitative research (objectivity in quantitative research) is the independence of findings from a particular investigator. We applied a form of investigator triangulation in that the second author reviewed and confirmed the first author's work along multiple dimensions: The second author both reviewed and confirmed (a) the search queries, selection filters, and their results and (b) the qualitative data analysis of the first author. In addition, working from a corresponding work log (laboratory book), the second author audited the steps taken by the first author to arrive at these findings (confirmability audit) and confirmed the findings.


\subsection{More limitations}
In addition to the qualitative research criteria discussed above, our study has several specific limitations typical of systematic literature reviews. Our focus on English-language publications may have introduced a language bias, potentially excluding relevant studies published in other languages. While the second author provided supervision and feedback throughout the coding process, we acknowledge that a more formalized inter-coder reliability assessment could have further strengthened our methodological rigor. However, the thorough review process and regular discussions between authors helped ensure consistency in data extraction and analysis.

A key limitation of our study is the relatively limited number of studies specifically focused on data engineering mistakes. To address this limitation, we deliberately included broader programming studies that provided insights into mistakes made with languages commonly used in data engineering (such as SQL for data manipulation and Java for data processing frameworks). This methodological choice allowed us to build a comprehensive foundation of programming mistakes while identifying patterns across technologies relevant to data engineering practice. However, many of our findings are derived from more general programming or data science contexts and may not fully capture the unique challenges in data engineering. This limitation is compounded by the fact that most of the studies we reviewed focused on beginner programming courses in academic settings. It's not clear how well these findings translate to beginners or novices in professional data engineering contexts.

The review may also be affected by publication bias, as studies with positive or significant results are more likely to be published than those with negative or non-significant results.

Furthermore, data engineering is a rapidly evolving field, and some of the older studies may not reflect the current state of tools and practices in the industry. Our study may also not fully capture all aspects of data engineering, as the field encompasses a wide range of skills and tools beyond just programming.

These limitations highlight the need for future research that focuses more specifically on data engineering contexts, particularly in professional settings, to validate and expand upon our findings.

\section{Conclusion}
This study highlights common programming mistakes made by beginners in data engineering, revealing challenges that range from basic syntax mistakes to complex, domain-specific issues. While general programming mistakes are widespread, the most significant hurdles for novices are unique to data engineering: misunderstanding datasets, inefficient algorithm implementation, and misinterpretation of data-specific problems.

Our findings have important implications not only for the development of data engineering tools but also for educational practices in the field. These insights can guide the creation of more intuitive programming languages and libraries specifically tailored to data engineering tasks, potentially mitigating common mistakes through improved design and functionality. Simultaneously, they underscore the need for educational interventions that focus on practical skill development and problem-solving abilities in real-world data engineering scenarios.

Future research could explore how addressing these common mistakes through educational methods impacts the efficiency and effectiveness of novice data engineers in both academic and real-world projects. Additionally, investigating the interplay between tool development, educational platforms, and practical experience could provide valuable insights for refining both technical tools and teaching methodologies.

This study serves as a foundation for improving not only the toolset available to aspiring data engineers but also the educational frameworks that support their growth. Ultimately, this dual approach will contribute to the development of more proficient professionals capable of addressing the complex data challenges of tomorrow, fostering a generation of data engineers who are better prepared for both technical and practical demands.


\bibliographystyle{IEEEtran}
\bibliography{references}

\appendix
\section{Literature} \label{appendix:literature}

\small The search process led to the following list of articles used in the literature review:

\begin{itemize}
    \small
    \item Singh et al., \textit{"Investigating Student Mistakes in Introductory Data Science Programming"} \cite{singh2024investigating}
    \item Brown and Altadmri, \textit{"Novice Java programming mistakes: Large-scale data vs. educator beliefs"} \cite{brown2017novice}
    \item Brown and Altadmri, \textit{"Investigating novice programming mistakes: Educator beliefs vs. student data"} \cite{brown2014investigating}
    \item Jegede et al., \textit{"Analysis of Syntactic Errors of Novice Python Programmers in a Nigeria University"} \cite{jegede2023analysis}
    \item Kaczorowska, \textit{"Analysis of typical programming mistakes made by first and second year IT students"} \cite{kaczorowska2020analysis}
    \item Smith and Rixner, \textit{"The error landscape: Characterizing the mistakes of novice programmers"} \cite{smith2019error}
    \item Altadmri and Brown, \textit{"37 million compilations: Investigating novice programming mistakes in large-scale student data"} \cite{altadmri201537}
    \item Hristova et al., \textit{"Identifying and correcting Java programming errors for introductory computer science students"} \cite{hristova2003identifying}
    \item Júnior et al., \textit{"Analyzing the Impact of Programming Mistakes on Students' Programming Abilities"} \cite{junior2019analyzing}
    \item Denny et al., \textit{"All syntax errors are not equal"} \cite{denny2012all}
    \item Ettles et al., \textit{"Common logic errors made by novice programmers"} \cite{ettles2018common}
    \item McCall and Kölling, \textit{"Meaningful categorisation of novice programmer errors"} \cite{mccall2014meaningful}
    \item Albrecht and Grabowski, \textit{"Sometimes it's just sloppiness-studying students' programming errors and misconceptions"} \cite{albrecht2020sometimes}
    \item Kaczmarczyk et al., \textit{"Identifying student misconceptions of programming"} \cite{kaczmarczyk2010identifying}
    \item Yarygina, \textit{"Learning analytics of CS0 students programming errors: The case of data science minor"} \cite{yarygina2020learning}
    \item Ahmadzadeh et al., \textit{"An analysis of patterns of debugging among novice computer science students"} \cite{ahmadzadeh2005analysis}
    \item Kiran and Moudgalya, \textit{"Evaluation of programming competency using student error patterns"} \cite{kiran2015evaluation}
    \item Jackson et al., \textit{"Identifying top Java errors for novice programmers"} \cite{jackson2005identifying}
    \item McCall and Kölling, \textit{"A new look at novice programmer errors"} \cite{mccall2019new}
    \item Miedema et al., \textit{"Identifying SQL misconceptions of novices: Findings from a think-aloud study"} \cite{miedema2022identifying}
    \item Vee et al., \textit{"Empirical study of novice errors and error paths in object-oriented programming"} \cite{vee2006empirical}
\end{itemize}

\section{Classification to Literature Mapping} \label{appendix:mapping}

\small This part of the appendix provides the mapping between each identified programming mistake type and the corresponding literature sources where these mistakes were observed and analyzed.

\begin{itemize}
    \small
    \item \textbf{Domain-Specific Mistakes (1)}
    \begin{itemize}
        \item (1.1) Dataset Misunderstandings \cite{singh2024investigating}
        \item (1.2) Faulty Data Analysis Logic \cite{singh2024investigating, yarygina2020learning}
        \item (1.3) Incorrect Data Handling \cite{singh2024investigating, yarygina2020learning}
        \item (1.4) Misunderstanding Data Types \cite{singh2024investigating, yarygina2020learning}
    \end{itemize}
    \item \textbf{Strategic Errors (2)}
    \begin{itemize}
        \item (2.1) Suboptimal Coding \cite{albrecht2020sometimes, singh2024investigating, vee2006empirical}
        \item (2.2) Wrong Algorithm \cite{ettles2018common, albrecht2020sometimes, singh2024investigating, kaczorowska2020analysis, vee2006empirical}
    \end{itemize}
    \item \textbf{Misinterpretation Errors (3)}
    \begin{itemize}
        \item (3.1) Incorrect Programming Assumptions \cite{denny2012all, ettles2018common, albrecht2020sometimes, brown2014investigating, kaczorowska2020analysis, altadmri201537, hristova2003identifying, ahmadzadeh2005analysis, jackson2005identifying, miedema2022identifying}
        \item (3.2) Task Misunderstanding \cite{ettles2018common, singh2024investigating, albrecht2020sometimes}
    \end{itemize}
    \item \textbf{Environment and Language Misconceptions (4)}
    \begin{itemize}
        \item (4.1) Incorrect Language Understanding \cite{ettles2018common, albrecht2020sometimes, singh2024investigating, kaczmarczyk2010identifying, yarygina2020learning, brown2017novice, brown2014investigating, jegede2023analysis, kaczorowska2020analysis, smith2019error, altadmri201537, hristova2003identifying, ahmadzadeh2005analysis, vee2006empirical, kiran2015evaluation, jackson2005identifying, mccall2019new, miedema2022identifying}
        \item (4.2) Library Misuse \cite{singh2024investigating, kaczorowska2020analysis, smith2019error, jackson2005identifying}
        \item (4.3) Path and I/O Errors \cite{albrecht2020sometimes, singh2024investigating, yarygina2020learning, kiran2015evaluation}
    \end{itemize}
    \item \textbf{Logical Errors (5)}
    \begin{itemize}
        \item (5.1) Incorrect Loop Conditions \cite{albrecht2020sometimes, kaczmarczyk2010identifying}
        \item (5.2) Faulty Conditional Logic \cite{ettles2018common, albrecht2020sometimes, kaczorowska2020analysis}
        \item (5.3) Off-by-One / Out-of-Bounds Errors \cite{ettles2018common, albrecht2020sometimes, kaczmarczyk2010identifying, kaczorowska2020analysis, smith2019error, ahmadzadeh2005analysis}
    \end{itemize}
    \item \textbf{Semantic Errors (6)}
    \begin{itemize}
        \item (6.1) Incorrect Function Usage \cite{singh2024investigating, yarygina2020learning, brown2014investigating, brown2017novice, kaczorowska2020analysis, hristova2003identifying, ahmadzadeh2005analysis, vee2006empirical, kiran2015evaluation, jackson2005identifying, mccall2019new, miedema2022identifying}
        \item (6.2) Type Mismatches \cite{denny2012all, ettles2018common, mccall2014meaningful, albrecht2020sometimes, brown2014investigating, brown2017novice, kaczorowska2020analysis, smith2019error, altadmri201537, hristova2003identifying, ahmadzadeh2005analysis, vee2006empirical, kiran2015evaluation, jackson2005identifying, mccall2019new}
        \item (6.3) Variable Scope Issues \cite{albrecht2020sometimes, singh2024investigating, kaczorowska2020analysis, kiran2015evaluation, miedema2022identifying}
    \end{itemize}
    \item \textbf{Syntax Errors (7)}
    \begin{itemize}
        \item (7.1) Incorrect Loop\&if-else Syntax \cite{brown2014investigating, kaczorowska2020analysis, altadmri201537, hristova2003identifying, jackson2005identifying}
        \item (7.2) Incorrect Operators \cite{ettles2018common, albrecht2020sometimes, singh2024investigating, brown2014investigating, brown2017novice, kaczorowska2020analysis, altadmri201537, hristova2003identifying, vee2006empirical, kiran2015evaluation, jackson2005identifying, miedema2022identifying}
        \item (7.3) Invalid Keywords Usage \cite{brown2014investigating, altadmri201537, hristova2003identifying}
        \item (7.4) Missing Semicolons \cite{denny2012all, mccall2014meaningful, albrecht2020sometimes, kaczorowska2020analysis, kiran2015evaluation, jackson2005identifying, mccall2019new}
        \item (7.5) Unbalanced Delimiters \cite{mccall2014meaningful, albrecht2020sometimes, brown2017novice, brown2014investigating, jegede2023analysis, kaczorowska2020analysis, altadmri201537, hristova2003identifying, jackson2005identifying, mccall2019new, miedema2022identifying}
        \item (7.6) Variable not declared or initialized \cite{ettles2018common, mccall2014meaningful, albrecht2020sometimes, singh2024investigating, jegede2023analysis, kaczorowska2020analysis, ahmadzadeh2005analysis, jackson2005identifying, mccall2019new}
    \end{itemize}
    \item \textbf{Sloppiness Errors (8)}
    \begin{itemize}
        \item (8.1) Typographical Errors \cite{mccall2014meaningful, albrecht2020sometimes, yarygina2020learning, vee2006empirical, jackson2005identifying, mccall2019new, miedema2022identifying}
    \end{itemize}
    \item \textbf{Memory Management Mistakes (9)}
    \begin{itemize}
        \item (9.1) Memory Language Misconceptions \cite{kaczmarczyk2010identifying, kaczorowska2020analysis, smith2019error}
        \item (9.2) Memory Leaks \cite{kaczorowska2020analysis, smith2019error}
    \end{itemize}
\end{itemize}

\end{multicols}

\end{document}